\documentclass[12pt,preprint]{aastex}

\shorttitle{SDSS QUASAR LENS SEARCH. V.}
\shortauthors{INADA ET AL.}

\begin{document}

\title{The Sloan Digital Sky Survey Quasar Lens Search. V. \\
Final Catalog from the Seventh Data Release}

\author{
Naohisa Inada,\altaffilmark{1,2} 
Masamune Oguri,\altaffilmark{3,4} 
Min-Su Shin,\altaffilmark{5,6} 
Issha Kayo,\altaffilmark{3,7} 
Michael A. Strauss,\altaffilmark{6} 
Tomoki Morokuma,\altaffilmark{8} 
Cristian E. Rusu,\altaffilmark{9,10}
Masataka Fukugita,\altaffilmark{3,11}
Christopher S. Kochanek,\altaffilmark{12} 
Gordon T. Richards,\altaffilmark{13}
Donald P. Schneider,\altaffilmark{14,15} 
Donald G. York,\altaffilmark{16,17}
Neta A. Bahcall,\altaffilmark{6}
Joshua A. Frieman,\altaffilmark{16,18,19}
Patrick B. Hall,\altaffilmark{20} and
Richard L. White\altaffilmark{21}
}

\altaffiltext{1}{Department of Physics, Nara National College of
  Technology, Yamatokohriyama, Nara 639-1080, Japan.}    
\altaffiltext{2}{Research Center for the Early Universe, School of
  Science, University of Tokyo, Bunkyo-ku, Tokyo 113-0033, Japan.}
\altaffiltext{3}{Institute for the Physics and Mathematics of the
  Universe, The University of Tokyo, 5-1-5 Kashiwa-no-ha, Kashiwa, 
  Chiba 277-8568, Japan.}
\altaffiltext{4}{Division of Theoretical Astronomy, National Astronomical 
  Observatory, 2-21-1, Osawa, Mitaka, Tokyo 181-8588, Japan.} 
\altaffiltext{5}{Department of Astronomy, University of Michigan, 
  500 Church Street, Ann Arbor, MI 48109-1042 USA.}
\altaffiltext{6}{Princeton University Observatory, Peyton Hall,
  Princeton, NJ 08544, USA.}                                   
\altaffiltext{7}{Department of Physics, Toho University, Funabashi,
  Chiba 274-8510, Japan.} 
\altaffiltext{8}{Institute of Astronomy, School of Science, University
  of Tokyo, 2-21-1 Osawa, Mitaka, Tokyo 181-0015, Japan.}
\altaffiltext{9}{Optical and Infrared Astronomy Division, National
  Astronomical Observatory of Japan, 2-21-1 Osawa, Mitaka, Tokyo
  181-8588, Japan.} 
\altaffiltext{10}{Department of Astronomy, Graduate School of Science,
  University of Tokyo 7-3-1, Hongo Bunkyo-ku, Tokyo 113-0033, Japan.}
\altaffiltext{11}{Institute for Cosmic Ray Research, University of Tokyo, 
  Kashiwa, 277-8582, Japan.} 
\altaffiltext{12}{Department of Astronomy, The Ohio State University, 
                  Columbus, OH 43210, USA.}                 
\altaffiltext{13}{Department of Physics, Drexel University, 3141
  Chestnut Street,  Philadelphia, PA 19104, USA.}
\altaffiltext{14}{Department of Astronomy and Astrophysics, The
  Pennsylvania State University, 525 Davey Laboratory, 
  University Park, PA 16802, USA.}   
\altaffiltext{15}{Institute for Gravitation and the Cosmos,The
  Pennsylvania State University, 525 Davey Laboratory, 
  University Park, PA 16802, USA.}   
\altaffiltext{16}{Department of Astronomy and Astrophysics, The University 
  of Chicago, 5640 South Ellis Avenue, Chicago, IL 60637, USA.}
\altaffiltext{17}{Enrico Fermi Institute, The University of Chicago,
  5640 South Ellis Avenue, Chicago, IL 60637, USA.} 
\altaffiltext{18}{Kavli Institute for Cosmological Physics, University 
  of Chicago, Chicago, IL 60637, USA.}
\altaffiltext{19}{Center for Particle Astrophysics, Fermi National 
  Accelerator Laboratory, P.O. Box 500, Batavia, IL 60510, USA.}
\altaffiltext{20}{Department of Physics and Astronomy, York University,
  4700 Keele Street, Toronto, Ontario, M3J 1P3, Canada.}
\altaffiltext{21}{Space Telescope Science Institute, 3700 San Martin
  Drive, Baltimore, MD 21218, USA.} 

\begin{abstract}
We present the final statistical sample of lensed quasars from the
Sloan Digital Sky Survey (SDSS) Quasar Lens Search (SQLS). 
The well-defined statistical lens sample consists of 26 lensed quasars
brighter than $i=19.1$ and in the redshift range of $0.6<z<2.2$ 
selected from 50,836 spectroscopically confirmed quasars in the SDSS
Data Release 7 (DR7), where we restrict the image separation range to
$1''<\theta<20''$ and the $i$-band magnitude differences in two image
lenses to be smaller than 1.25~mag. The SDSS DR7 quasar catalog also
contains 36 additional lenses identified with various techniques. In
addition to these lensed quasars, we have identified 81 pairs of
quasars from follow-up spectroscopy, 26 of which are physically
associated binary quasars. The statistical lens sample 
covers a wide range of image separations, redshifts, and magnitudes,
and therefore is suitable for systematic studies of cosmological
parameters and surveys of the structure and evolution of galaxies and
quasars.   
\end{abstract}

\keywords{gravitational lensing: strong --- quasars: general --- 
          cosmology: observations}

\section{INTRODUCTION}\label{sec:introduction}

The Sloan Digital Sky Survey Quasar Lens Search
\citep[SQLS;][]{oguri06a,oguri08a,inada08a,inada10}  is a large
survey for gravitationally lensed quasars among the spectroscopically 
confirmed quasars in the Sloan Digital Sky Survey \citep[SDSS;][]{york00}. 
The SQLS is designed for statistical analyses, and hence pays
particular attention to the selection function of the lens sample by
taking advantage of the homogeneous SDSS data set. For rigorous
statistical analyses, the so-called ``statistical sample'' is designed
to be complete within its prescribed limits.  
Well-defined statistical lens samples can be used for
a number of cosmological and astrophysical applications. For instance,
the number counts of strong lenses in a given source quasar population
is a useful constraint on the cosmological constant
\citep{turner90,fukugita90}. In addition, one can study the evolution
of massive galaxies using the statistics of strong lenses, including
the redshift distribution of lens galaxies 
\citep{kochanek92,ofek03,rusin03,rusin05,koopmans06,koopmans09,auger10,faure11}.  

Statistical lens samples have been constructed from both radio and
optical quasar surveys. For instance, the Hubble Space Telescope 
({\it HST}) Snapshot survey \citep{maoz93} includes five lenses
discovered among 502 bright high-redshift quasars. The Cosmic-Lens
All Sky Survey \citep[CLASS;][]{myers03,browne03} was the largest
quasar lens survey before the SQLS, and contains 13 lenses identified 
from a sample of 8958 flat spectrum radio sources for the 
statistically well-defined sample. However, a large source of the
systematic error in the statistical analysis of the CLASS lens sample
comes from the poorly characterized redshift distribution of the
sources \citep[e.g.,][]{munoz03}. 

In addition to strongly lensed quasars, strongly lensed galaxies are
also searched for. In particular, recent wide-field
urveys have produced large samples of galaxy-galaxy lenses
\citep[e.g.,][]{bolton08a,faure09,marshall09,feron09,kubo10,treu11,more12,brownstein12},
which are principally useful for studies of the lens galaxies.
Quasar lenses are more useful as cosmological probes because 
they represent a better understood source population and it is
easier to characterize the selection effects of point
sources. Moreover, time delays between multiple images of quasars 
provide additional information on cosmological parameters and the
structure of the lens galaxies 
\citep[e.g.,][]{kochanek02,saha06,oguri07,keeton09,suyu10,paraficz10,courbin11}.

In \citet{inada10}, we presented the SQLS statistical lens sample
of 19 lenses from the full SDSS-I data set \citep[DR5;][]{adelman07}.
In this paper, we present the final result of the SQLS from SDSS 
Data Release 7 \citep[DR7;][]{abazajian09}. DR7 is the final data
release of SDSS-II, and consists of about 8000~deg$^2$ of
spectroscopic coverage. We select lens candidates from the DR7
spectroscopic quasar catalog of 105,783 quasars from
\citet{schneider10}. The selection process is the same as
that used for the DR3 and DR5 statistical samples 
\citep{inada08a,inada10}. As in previous SQLS papers, 
we present the results of confirming observations for all the
candidates, including quasars that were shown not to be gravitational
lenses. Several of the candidates are found to be binary quasars
(physical pairs of quasars with small separations). 

In this paper, we describe our source
quasar sample in Section~\ref{sec:source}, and identify lens candidates
in Section~\ref{sec:candidate}. Follow-up observations of the lens
candidates are presented in Section~\ref{sec:observation}. 
Section~\ref{sec:lens_catalog} discusses the lens sample, and we
summarize our results in Section~\ref{sec:summary}. When necessary, we
assume a flat universe with matter density $\Omega_M=0.275$, 
cosmological constant $\Omega_\Lambda=0.725$, and the Hubble constant
$h=H_0/(100{\rm km\,s^{-1}Mpc^{-1}})=0.702$ \citep{komatsu11}. 

\section{SOURCE QUASARS}\label{sec:source}

The SDSS consists of imaging and spectroscopic surveys of a quarter of
the sky using a dedicated wide-field 2.5-meter telescope
\citep{gunn06} at the Apache Point Observatory in New Mexico, USA.  
The imaging survey uses five broad-band filters
\citep[$ugriz$,][]{fukugita96,gunn98,doi10}. The automated pipeline
leads to astrometric accuracy better than about $0\farcs1$
\citep{pier03}, and photometric zeropoint accuracies of about 0.01
magnitude \citep{hogg01,smith02,ivezic04,tucker06,padmanabhan08}.  
The spectroscopic targets  selected based on the imaging data,
including candidate quasars selected based on colors 
\citep{richards02}, are observed with multi-object fiber spectrographs
covering 3800{\,\AA} to 9200{\,\AA} with a resolution of $R\sim1800$
\citep{stoughton02,blanton03}.
All the SDSS-I and SDSS-II data are publicly available 
\citep{stoughton02,abazajian03,abazajian04,abazajian05,abazajian09,
adelman06,adelman07,adelman08}. 

The source quasar sample is constructed following the procedures
detailed in \citet[][hereafter Paper I]{oguri06a}, \citet[][hereafter
  Paper II]{inada08a}, and \citet[][hereafter Paper IV]{inada10}.
An important difference from the DR5 study is that the DR7 quasar
catalog contains quasars found in a series of ``Special
Plates''\citep[see][]{adelman06,schneider10}. Most of the Special
Plates were obtained in observations along the celestial equator at
the Southern Galactic Cap, and used different selection algorithms
from the main survey to explore the boundaries of the selection
algorithms.  The remainder comes from Sloan Extension for the 
Galactic Understanding and Exploration \citep[SEGUE;][]{yanny09}. 
Clearly these quasars are not appropriate for our statistical lens
sample, because they have quite different selection functions and
redshift and magnitude distributions from the main survey. We do not 
include quasars from the Special Plates in the statistical lens sample.  

For the statistical sample, we start with the 105,783 quasars in the
DR7 spectroscopic survey selected over 8000~${\rm deg^2}$
\citep{schneider10}. As discussed above, we first remove 9,742 quasars
in the Special Plates, as they are not suited for the statistical
analysis. We then restrict the redshift range to $0.6<z<2.2$ and the
$i$-band Point Spread Function (PSF) magnitude to $i<19.1$ after
correcting for Galactic extinction based on
\citet{schlegel98}. The quasar target selection is almost complete
over these redshift and magnitude ranges (see \citealt{richards06};
Paper I). In addition, we exclude quasars in poor seeing fields ({\tt
  PSF\_WIDTH}$>1\farcs8$ in $i$-band) where selection of small-separation lenses
becomes very inefficient. These selection criteria lead to a sample of
50,836 source quasars.  We give the catalog of the source quasars in
Table~\ref{tab:source}.

\section{LENS CANDIDATES}\label{sec:candidate}

The SQLS uses two different algorithms for selecting lens 
candidates. ``Morphological selection'' is used to find smaller
separation ($\theta\lesssim 2\farcs5$) lensed quasars which are 
not deblended into multiple components by the SDSS photometric
pipeline. The other is ``color selection'' to find large
separation lenses ($\theta\gtrsim 2\farcs5$) whose lensed quasar
components are found as separate objects in the SDSS image catalog.  

The specific procedures of the algorithms and the selection 
functions are detailed in Papers I and II. In brief, morphological
selection identifies quasars that are extended and poorly fit by the
PSF. In morphological selection we check the parameter {\tt 
  star\_L}, a logarithmic likelihood that the object is fit by the
PSF, and the parameter {\tt objc\_type}, which assigns a
classification of the object as star or galaxy. The algorithm was
tested against many simulated lensed quasars in Paper I. We then  
fit the SDSS image of each system with a model consisting of two
stellar components using GALFIT \citep{peng02,peng10}, and exclude objects
that have large flux differences or have separations between
the two PSF components that are too small in both the $u$- and
$i$-band SDSS images. This procedure is necessary to reject obvious
single quasars and quasar-galaxy superpositions. We then visually
inspect the images of candidates that survive the GALFIT selection
step to exclude other obvious superpositions of a quasar and a
galaxy. Of the 50,836 DR7 source quasars, 187 lens candidates are
selected by the morphological selection criteria. Finally we compare
the candidates with DR3 (Paper II) and DR5 (Paper IV) candidates, and
find that 133 candidates have previously been examined. The remaining
54 morphological candidates are subject to detailed study in this
paper. 

We also identify lens candidates based on our color selection criteria
for cases where the lensed images are deblended by the
photometric pipeline. For each source quasar, we search for objects in
the vicinity that
have colors similar to those of the quasar.  We select objects with 
an angular separation $\theta<20\farcs1$ and $i$-band magnitude
differences $|{\Delta} i|<1.3$, in order to obtain a complete
sample with image separations $1''<\theta<20''$ and $i$-band magnitude
differences between two images $|{\Delta}i|<1.25$ with some
allowance. We then compare optical and radio flux ratios, which 
are measured from the Faint Images of the Radio Sky at Twenty 
centimeters survey \citep[FIRST;][]{becker95}, and exclude candidates
that show clearly inconsistent flux ratios between the optical and radio.  
Given the resolution of the FIRST data, we examine the radio data only
for candidates with image separations larger than $6''$. 
We also reject obvious quasar-galaxy pairs by visual inspection. 
Furthermore, we exclude low-redshift, large-separation pairs with
no detectable lensing objects in the SDSS images, because at least one
of the member galaxies of the putative lens group/cluster should be
detectable for such lensing events (see Paper II for quantitative
details). This leaves 340 color selected candidates out of 50,836 DR7
source quasars. Since 216 of these 340 candidates were previously
examined in Paper II and Paper IV, we investigate the remaining 124
color candidates in this paper.  

To summarize, we identified 54 and 124 new candidates based on our
morphological and color selection criteria, respectively, which
required additional observations to complete the construction of the
DR7 statistical lens sample. Two candidates are selected by both
methods, because their image separations of $\theta\sim 2''-3''$
correspond to the transition of the image separation ranges probed by
the two selection algorithms. Thus the total number of unique new
candidates is 176. Table~\ref{tab:process} summarizes how many lens
candidates are rejected at each step of the process.  

\section{OBSERVATIONS}\label{sec:observation}

\subsection{Summary of Follow-up Observations}\label{sec:followup}

Among the 176 new lensed quasar candidates selected in the previous
section, two systems are previously discovered lenses or binary
quasars. One is SDSS~J095122.57+263513.9 
\citep[FBQ0951+2635,][]{schechter98}, a gravitational lens with an
image separation of $\theta=1\farcs1$ discovered in the FIRST Bright
QSO Survey \citep{gregg96}. The other, SDSS~J143002.66+071415.6,
was shown to be a pair of quasars at $z=1.246$ and $z=1.261$ 
\citep{myers08}. It was discovered in the course of a binary quasar
survey of the SDSS photometric quasar catalog of \citet{richards04}.

We observed all the remaining 174 lens candidates using various
telescope facilities, which are summarized in Tables~\ref{tab:can_mor} 
and \ref{tab:can_col}.\footnote{We provide a list of all the 520 
targets, including those investigated in Paper II and Paper IV, and the
summaries of observations of these candidates on the SQLS webpage
{\tt http://www-utap.phys.s.u-tokyo.ac.jp/\~{}sdss/sqls/}.} 
These observations consist of optical 
imaging, near-infrared imaging, and optical spectroscopy. We follow
Paper II and Paper IV for the follow-up strategy, which is briefly
summarized here. For the morphological candidates, we usually start
with optical imaging under good seeing conditions (seeing FWHM
$\lesssim 1''$) to see if the candidates have two or more
stellar components and a lens galaxy between the stellar
components. Most of the candidates are easily rejected by such
observations, because they either have only a single component or show 
no sign of a lens galaxy. We carry out spectroscopy of multiple
stellar components for candidates which show plausible lens
galaxies in the follow-up images. We obtained spectra of 13
morphological candidates, 6 of which are true gravitational
lenses. Two additional candidates, SDSS~J125107.57+293540.5 and
SDSS~J133018.64+181032.1, were confirmed as gravitational lenses
without needing spectroscopy because of their obvious quadruple image
configurations.  
For the 122 color-selected candidates, excluding the two candidates
that were  also selected as morphological candidates, we discovered many
pairs of quasars in the spectroscopic follow-up observations (see
\S~\ref{sec:note} and \S~\ref{sec:binary}). Based on the similarity of
the spectra of the stellar components and the detection of the lensing
object, we conclude that only 2 of the 122 color candidates are true 
gravitational lens systems. To summarize, these follow-up observations
led to the discoveries of 10 gravitational lenses, which
are reported in \citet{kayo07}, \citet{kayo10}, \citet{oguri08d},
\citet{jackson12}, Rusu et al. (2012, in preparation), and Inada et
al. (2012, in preparation).

\subsection{Note on individual objects}\label{sec:note}
 
In addition to our confirmed gravitational lens systems, we found 24
objects that have multiple quasar components. Here we discuss the
three binary quasar pairs with velocity differences smaller than
$1000\,{\rm km\,s^{-1}}$ and a morphological candidate which is a
potential gravitational lens outside the statistical sample.  

{\bf SDSS~J094234.97+231031.1:}
Our spectroscopy at the 6.5-meter MMT indicates that both
stellar components, separated by $\theta=2\farcs5$, have similar spectra
of a quasar at $z=1.833$, but the shapes of the \ion{C}{4} and
\ion{C}{3]} emission lines are significantly different. 
We also find no evidence for a lens between them in images taken with
UH8k at the University of Hawaii 2.2-meter telescope ($R$ and
$I$-band) or with NICFPS on the Apache Point Observatory 3.5-meter
telescope ($K$-band). Therefore, we conclude that this object is a
binary quasar.    

{\bf SDSS~J112235.03+232634.9:}
This is a small separation ($\theta\sim 0\farcs7$) lens candidate
identified based on its morphology. We obtained spectra with
FOCAS \citep{kashikawa02} at the Subaru 8.2-meter telescope and 
$I$-band images with the University of Hawaii 2.2-meter telescope,
but these observations have been inconclusive. Because of the small 
image separation, this system would not be included in the statistical 
lens sample even if confirmed.

{\bf SDSS~J124257.32+254303.0:}
Spectra taken at the Subaru 8.2-meter telescope indicate that the two
stellar components separated by $\theta=2\farcs8$ are quasars at
slightly different redshifts of $z=0.827$ and $z=0.824$ ($\Delta v
\sim 560 {\rm km\,s^{-1}}$), as measured
from the \ion{Mg}{2} emission lines. Images taken with tek2k at the
University of Hawaii 2.2-meter telescope ($I$-band) show extended
emission  between the two components, but we interpret this as host
galaxy emission of one or both quasars, particularly given the low
quasar redshifts. 

{\bf SDSS~J143350.94+145008.1:}
While the two stellar components with a separation of $\theta=3\farcs3$
have similar redshifts of $z=1.506$, based on spectroscopy at the
Apache Point Observatory 3.5-meter telescope, the two spectra are 
markedly different. One component has broad absorption features on the
blue side of the \ion{C}{4} and \ion{C}{3]} emission lines, which are
not seen in the spectrum of the other component. In addition, imaging
at the University of Hawaii 2.2-meter telescope ($I$-band) shows no
sign of a lens galaxy. We conclude that this object is a binary
quasar.   

\section{LENSED QUASAR CATALOG}\label{sec:lens_catalog}

\subsection{Statistical Sample}\label{sec:statsample}

We define the statistical lens sample to include objects with an image
separation in the range of $1''<\theta<20''$ and an $i$-band magnitude
difference less than 1.25~mag for doubles, as was adopted in Paper II
and Paper IV.  
The ranges are determined such that the lens candidate selection is
almost complete (see Paper I). Among the 11 new lensed quasars since
DR5 selected by these algorithms, 7 lenses satisfy the
criteria. Together with 19 lenses from the DR5 sample (Paper IV), the
statistical lens sample consists of 26 lenses selected from a sample
of 50,836 source quasars. 

The properties of this statistical lens sample are summarized in 
Table~\ref{tab:lens_dr7stat}. Among the 26 lenses there are only 
four four-image lenses (SDSS~J092455.79+021924.9, PG1115+080,
SDSS~J125107.57+293540.5, and SDSS~J133018.64+181032.1) and one 
five-image lens, the cluster lens SDSS~J100434.92+411242.7
\citep{inada03b,oguri04b,inada05a}.  In addition to basic parameters
such as source redshifts and $i$-band magnitudes, we show the
magnitudes of the lens galaxies measured in the discovery papers or
by the {\it HST} from the CASTLES webpage if available. We also
revisited the lens redshifts for some of the lens systems. 
New spectroscopy with Gemini/GMOS allows us to
easily measure the lens redshift of SDSS~J100128.61+502756.8 to be
$z=0.415$ (N. Inada et al., in preparation). A re-examination of
the spectrum of SDSS~J105545.45+462839.4 \citep{kayo10} shows Mg
and Na absorption lines in the spectrum of the lens galaxy, from
which we determine the lens redshift to be $z=0.388$. We detected
\ion{Mg}{2} absorption lines in the quasar spectrum of
SDSS~J102111.01+491330.3 at a redshift of $z=0.451$ that agrees well
with the lens redshift expected from the color and brightness of the
lens galaxy.

\subsection{Additional Lensed Quasars}\label{sec:addlens}

As in previous SQLS data release papers (Paper II and Paper IV), 
we have identified several new lenses from the DR7 quasar catalog that
are not included in the statistical lens sample. 
Table~\ref{tab:lens_dr7add} summarizes these 36 additional
lenses, 18 of which are new quasars in the DR7 catalog that were not
include in the DR5 quasar catalog.
Many were discovered by the SQLS simply by applying our
selection algorithms to quasars that are outside the redshift,
magnitude, and image separation ranges of the statistical 
lens sample (see also Paper IV). The large-separation 
lens SDSS~J102913.94+262317.9 \citep{inada06b,oguri08c} is included 
in this category because its image separation of $\theta=22\farcs5$ is
slightly larger than the image separation limit for the statistical
lens sample. Several lenses were also discovered by examining the
morphology of SDSS quasars in the UKIDSS \citep[UKIRT Infrared Deep Sky
  Survey;][]{lawrence07}, by taking advantage of the higher image
  quality and the more dominant contribution of lens galaxies in
the near-infrared \citep{jackson08,jackson09,jackson12}. 

We can also update two of these lens redshifts. The lens redshift of
SDSS~J125819.24+165717.6 measured in our follow-up Gemini/GMOS
spectroscopy is $z=0.505$ (N. Inada et al., in preparation).
From a re-examination of the spectrum of SDSS~J130451.52+161138.2
\citep{kayo10}, we conclude that the lens redshift is likely to be
$z=0.373$.  

\subsection{Properties of the Lens Samples}\label{sec:property}

\begin{figure}
\epsscale{.47}
\plotone{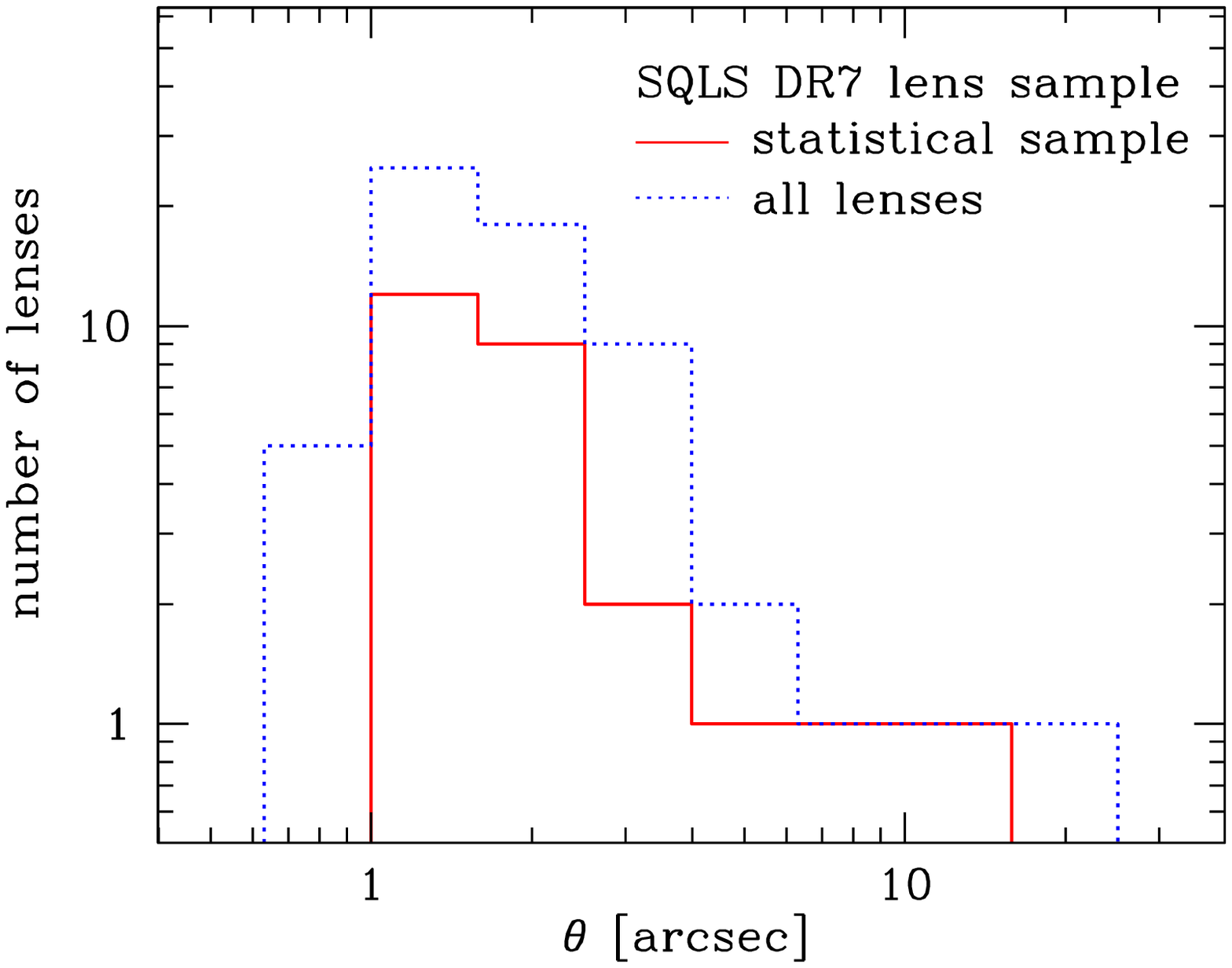}
\plotone{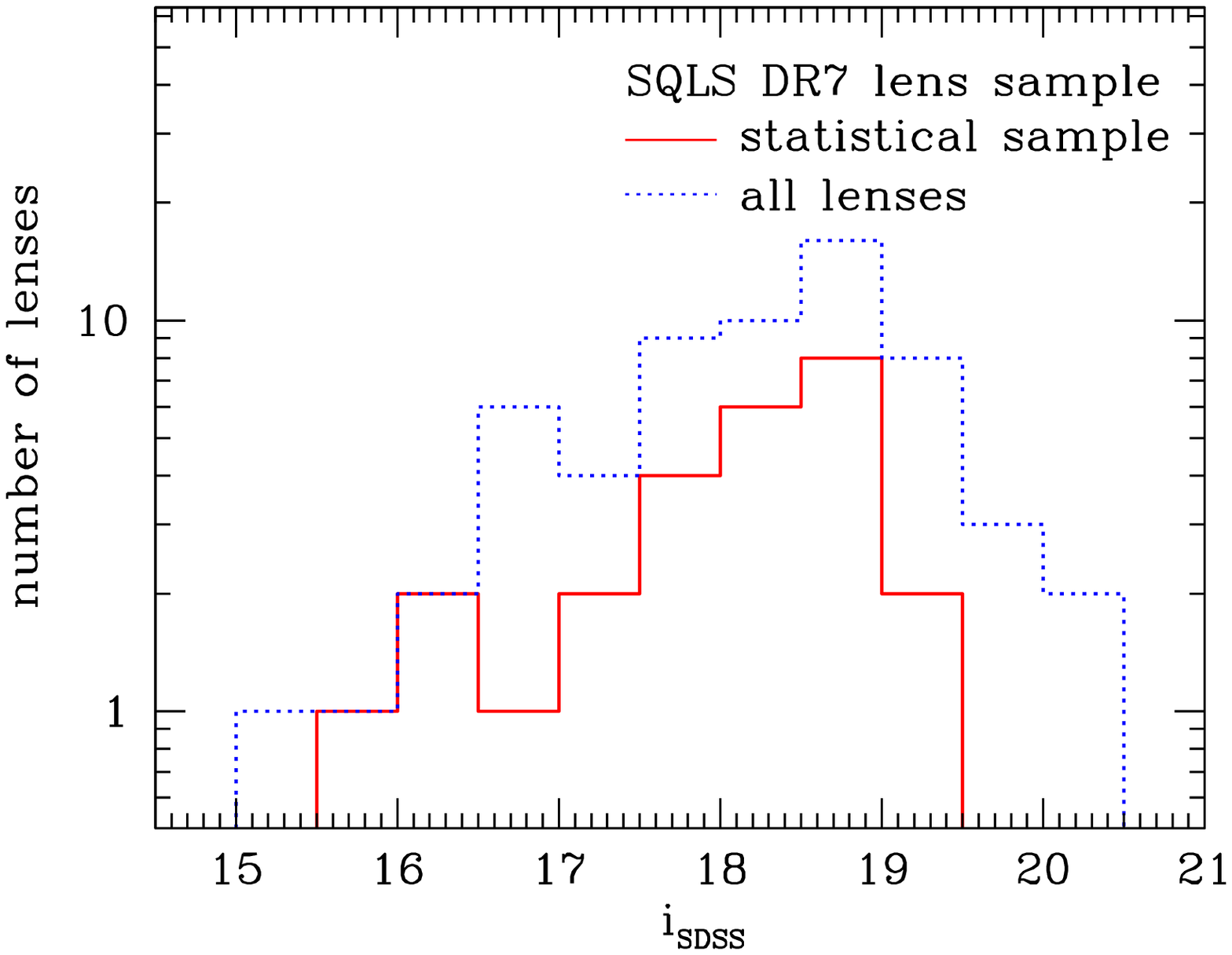} \\
\plotone{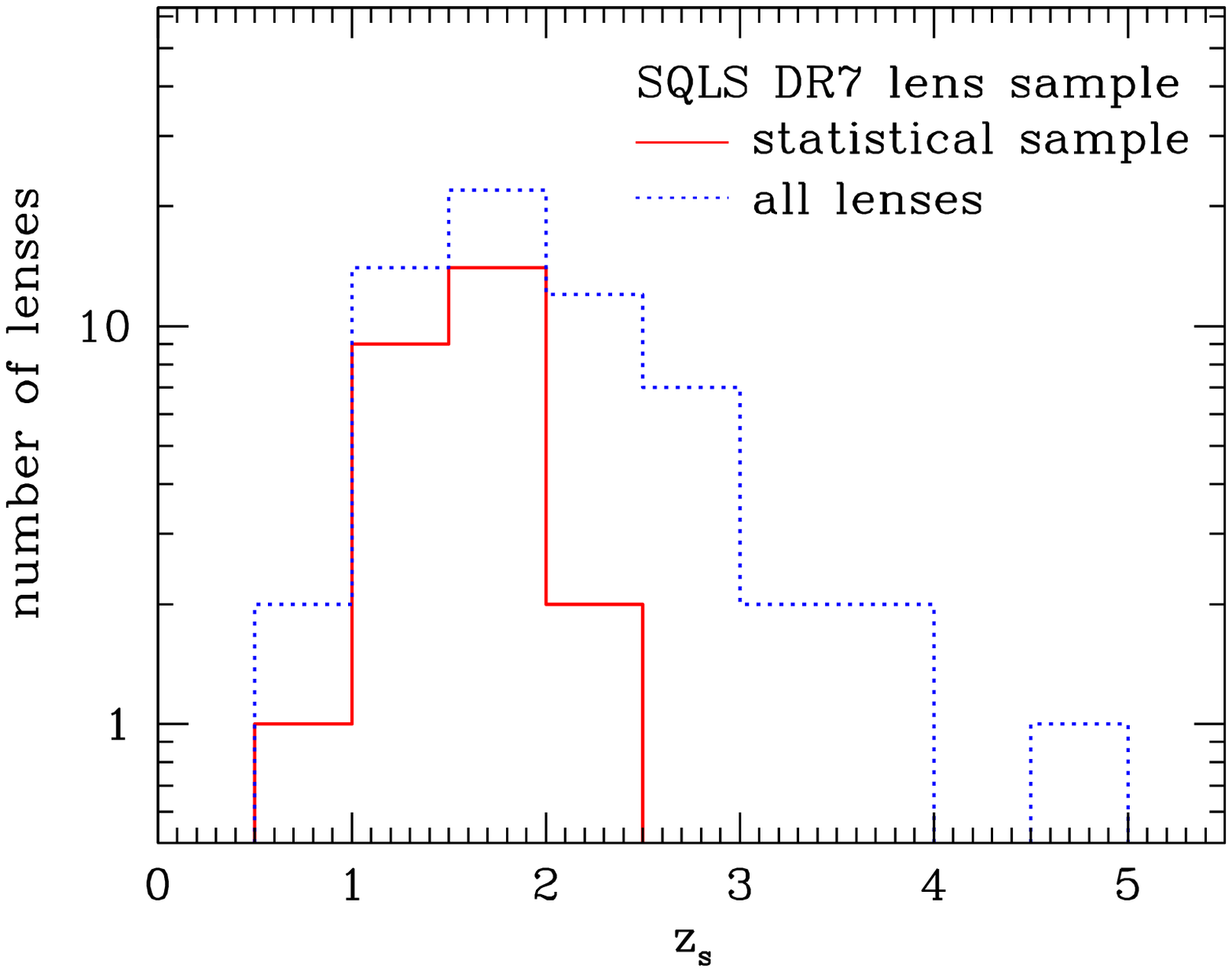}
\plotone{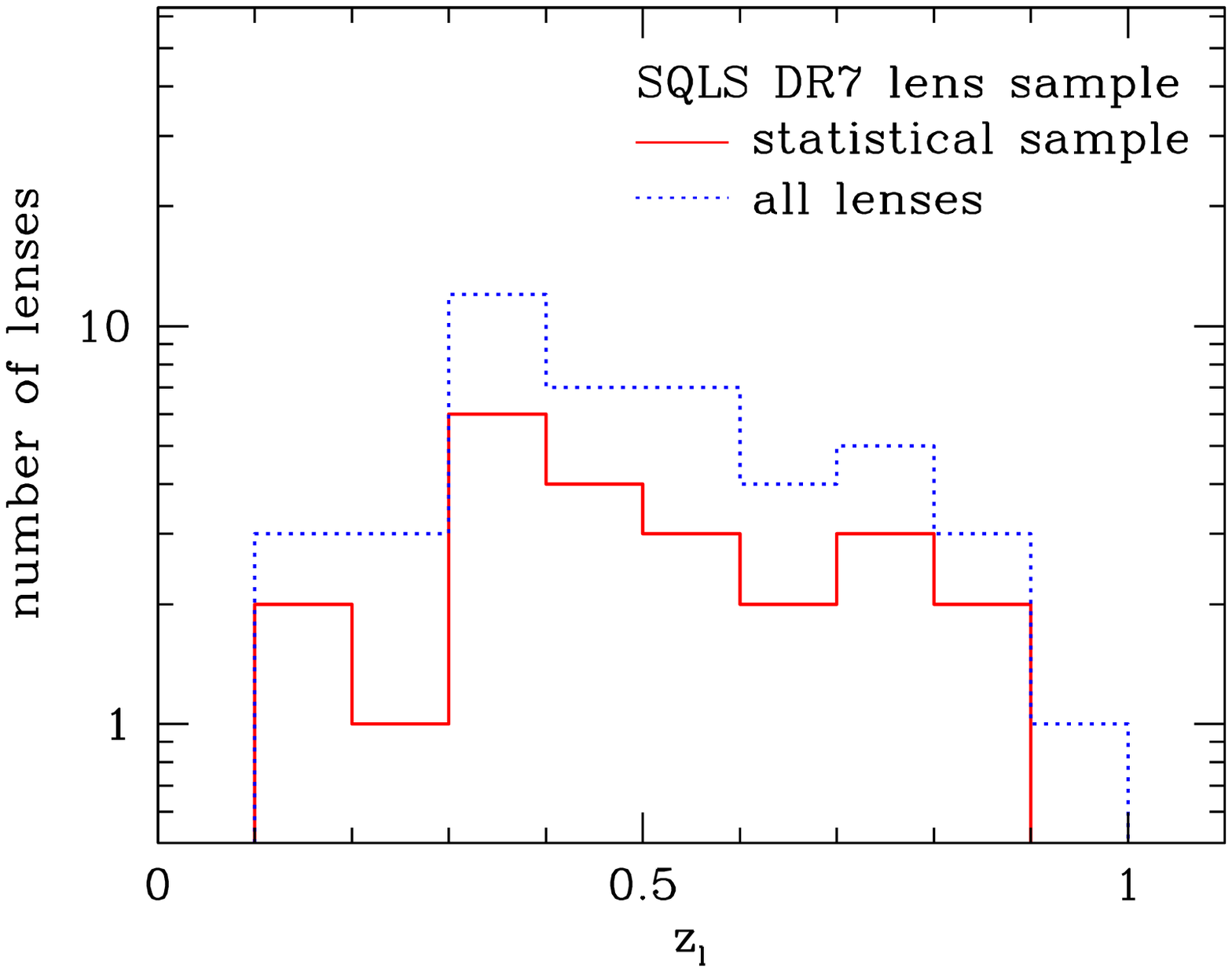}
\caption{Properties of the SQLS lenses in the statistical sample
  ({\it solid}) and all lenses ({\it dotted}). Distributions of the
  image separations $\theta$ ({\it upper left}), the $i$-band PSF magnitudes
  $i_{\rm SDSS}$ ({\it upper right}), the source redshifts
  $z_s$ ({\it lower left}), and the lens redshifts $z_l$ ({\it lower
    right}) are shown. For the lens redshift distribution, we use only
  the lens systems whose lens redshifts are measured
  spectroscopically (including those measured by absorption lines in
  the quasar spectra).
\label{fig:sep_dr7}} 
\end{figure}

In Figure~\ref{fig:sep_dr7}, we show the distributions of our lens
samples in several variables. The image separation distribution
clearly indicates that our sample covers a wide range of lens types,
ranging from lensing by single galaxies ($\theta\sim 1''$) to 
clusters of galaxies ($\theta\gtrsim 10''$). The distribution is in
reasonable agreement with that of strongly lensed galaxies in the CFHT
Legacy Survey \citep{more12} as well as with theoretical expectations
\citep[e.g.,][]{oguri06b}. Our lens sample contains many lensed
quasars with $i\sim 19$, close to the magnitude limit of SDSS
spectroscopic quasars in this redshift range. The average source
redshift of the statistical sample is $\langle z_s\rangle =1.56$,
whereas that of the entire sample is $\langle z_s\rangle =1.95$. The
lens redshifts are also distributed over a wide range, from  
$z_l\sim 0.2$ to $z_l\sim 1$.  

\begin{figure}
\epsscale{.55}
\plotone{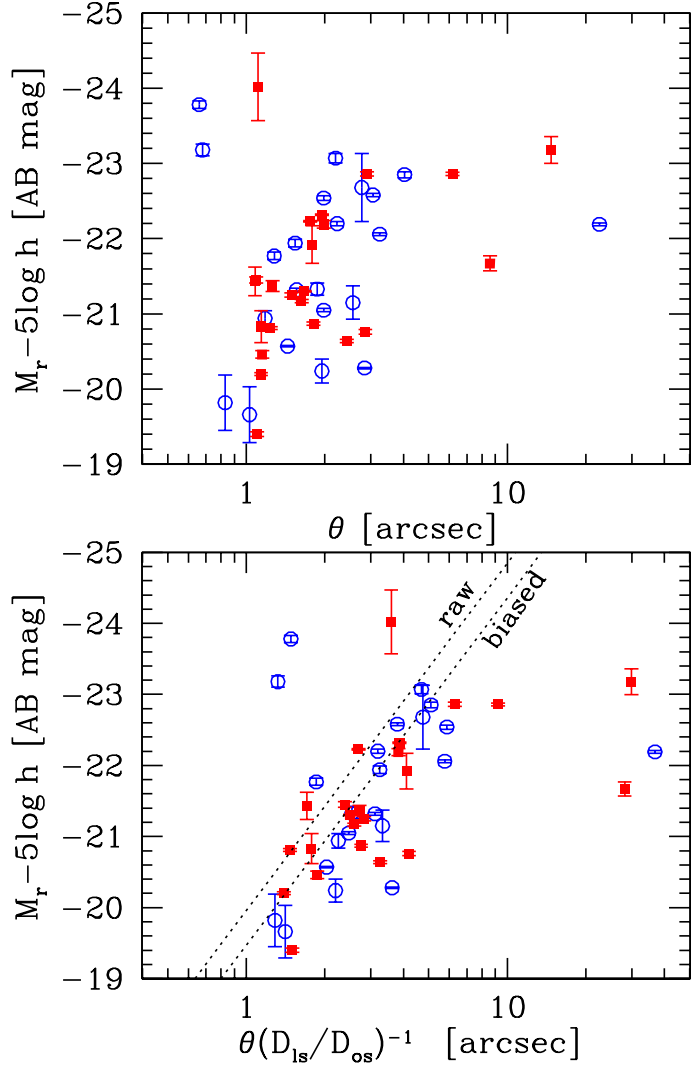}
\caption{The correlation between the image separation $\theta$ and
  the absolute $r$-band magnitude of the lens galaxy $M_r$. The
  absolute magnitudes are computed from the observed $I$-band
  magnitudes listed in Tables~\ref{tab:lens_dr7stat} and
  \ref{tab:lens_dr7add}. A template spectrum of an elliptical galaxy
  by \citet{coleman80} is used to compute the cross-filter
  $K$-correction. Filled squares show lenses in the statistical
  sample and open circles are additional lenses. Error-bars show only 
  the statistical errors of the apparent magnitude measurements.  
  While the raw image separations are used in the upper panel, in the
  lower panel we rescale the image separation with the distance ratio
  $(D_{ls}/D_{os})^{-1}$ in order to eliminate the redshift dependence
  of the image separation. The dotted lines show the predicted
  correlation from the Faber-Jackson relation measured in the SDSS
  data with the luminosity evolution extrapolated to $z=0.5$
  \citep{bernardi03} with and without the effect of the lensing
  selection bias (see text for details).
\label{fig:mabs_sep}} 
\end{figure}

The image separations are expected to be larger for more massive 
lensing objects, but with a sharp cutoff between galaxy and cluster
mass scales due to the effect of baryon cooling
\citep[e.g.,][]{kochanek01}. For the singular isothermal sphere model, 
the image separation is $\theta=8\pi(D_{ls}/D_{os})(\sigma_v/c)^2$,
where $D_{ls}$ and $D_{os}$ are the angular diameter distances between
the lens and the source and between the observer and the source,
respectively. The well-known scaling relation between velocity
dispersions $\sigma_v$ 
and galaxy luminosities \citep{faber76} suggests that the image
separations should correlate  with galaxy luminosities. The
correlation between  the image separation $\theta$ and the $r$-band  
absolute magnitude of the lens galaxy is plotted in
Figure~\ref{fig:mabs_sep}. The absolute magnitudes are computed from
apparent magnitudes of lens galaxies listed in
Tables~\ref{tab:lens_dr7stat} and \ref{tab:lens_dr7add} using the
cross-filter $K$-correction from a template spectrum of an elliptical
galaxy by \citet{coleman80}. 
We see a clear correlation between $\theta$ and $M_r$, such that
lenses with larger image separations have more luminous lens
galaxies, consistent with the expectation and earlier lensing results 
\citep[e.g.,][]{rusin03,bolton08b}. Independently of the mass model,
the image separation scales with the distance ratio $D_{ls}/D_{os}$,
and thus we also consider the scaled image separation
$\theta(D_{ls}/D_{os})^{-1}$ to remove the dependence on a
distance. As expected, the scaled image separations show a tighter 
correlation with the absolute magnitudes. A few lenses
with $\theta\gtrsim 10''$ are offset from the correlation.
These are cluster lenses where there is a much larger contribution of
dark matter inside the Einstein radius
\citep[e.g.,][]{kochanek01}. We note that two previously known
subarcsecond lenses PMN~J0134$-$0931 and FBQ0951+2635 show large
deviations from the correlation. One of the reasons is that their lens
galaxies are likely to be spiral galaxies for which the Faber-Jackson 
relation cannot be applied, but another reason might be due to errors
in magnitudes or redshifts estimates for their lens galaxies. 

The distance-scaled image separation
$\theta(D_{ls}/D_{os})^{-1}$ should be independent
of redshift for the standard  singular isothermal sphere model. Hence
by adopting the Faber-Jackson relation \citep{faber76}, the
correlation between the velocity dispersion $\sigma_v$ 
and the absolute magnitude of early-type galaxies, we can predict
the expected $\theta(D_{ls}/D_{os})^{-1}$-$M_r$ correlation. Note that
most of lens galaxies in our sample are early-type galaxies.
In Figure~\ref{fig:mabs_sep}  we also show the correlation expected from
the Faber-Jackson relation observed in the SDSS data \citep{bernardi03}.
Here we include luminosity evolution by extrapolating the redshift
evolution measured by \citet{bernardi03} to $z=0.5$, which is roughly
the median lens redshift for our sample (see Figure~\ref{fig:sep_dr7}). 

We find that the predicted relation properly reproduces the observed
slope, except for the largest separation lenses, where the lens
galaxies are much fainter than the relation predicts due to the
dominant contribution of dark matter to these image separations. On
the other hand, the normalization is slightly offset, in the sense
that the observed lens galaxies are on 
average fainter than the relation predicts. One cause of this shift is
a bias created by the large scatter of the Faber-Jackson
relation \citep{kochanek94}. \citet{bernardi03} found a scatter in 
$\log \sigma_v$ of $\sim 0.07$ at fixed galaxy luminosity. 
Since the lensing cross section is proportional to $\sigma_v^4$, the
lensing-weighted average of the velocity dispersion is given by the
average over the Gaussian in $\log \sigma_v$ times $\sigma_v^4$. 
Thus the lensing bias shifts the mean $\log \sigma_v$ by  $\Delta
(\log \sigma_v)\sim 0.05$, which roughly corresponds to the image
separation shift of $\Delta(\log \theta)=2\Delta (\log \sigma_v)\sim
0.1$. Including this lensing bias in the predicted correlation agrees
much better with observations, as shown in
Figure~\ref{fig:mabs_sep}. This comparison highlights the importance
of understanding the selection effects for studying galaxy properties
from observed strong lens systems.  

\subsection{Binary Quasars and Projected Quasar Pairs}\label{sec:binary}

One important by-product of the SQLS is the discovery of many quasar
pairs and binary quasars. In total we have identified 81 pairs of
quasars, 26 of which are quasar pairs with velocity differences
smaller than the $2000~{\rm km\,s^{-1}}$ threshold used for the
statistical analysis of binary quasars in \citet{hennawi06a}. Most of
these binary quasars and projected quasar pairs are new. In
particular, we have identified 12 pairs of quasars with angular
separations smaller than $3''$. Detailed explorations of these 
small separation binary quasars are important for understanding the
possible connection of quasar activity with galaxy mergers
\citep[e.g.,][]{green10}. These quasar pairs also provide
useful information about the nature of quasars by constraining their
small-scale correlation function \citep{hennawi06a,myers08} and the
distribution of absorbers around quasars
\citep{bowen06,hennawi06b,tytler09}.    

\section{SUMMARY}\label{sec:summary}

We have presented the final statistical lens sample of the SQLS from
the SDSS DR7 quasar catalog \citep{schneider10}. We observed 176
lensed quasar candidates selected from 50,836 quasars with various
facilities to determine whether they are in fact lenses.  
Together with the results we reported in Paper II and Paper IV, we
have constructed a sample of 26 lensed quasars within a well-specified
separation range and magnitude difference. The well-defined
statistical sample presented in the paper will be essential for
studying dark energy as well as the evolution of massive clusters. We
present a detailed statistical analysis of this final lens sample in a
separate paper \citep[][paper VI]{oguri12}. 
In addition to the statistical sample, the DR7 quasar catalog contains
at least 36  additional lenses identified with various techniques. 
These additional lenses will also be useful for detailed future
studies, including precise mass modeling from high-resolution
astrometry, time delay measurements, the measurement of the lens
velocity dispersion, and the study of high-redshift quasars through host
galaxy or microlensing observations. The SQLS has also discovered many
small-separation binary quasars and projected quasar pairs, which may
provide a clue to the origin and nature of quasar activity.

The largest quasar lens survey before SQLS was CLASS,
which contains 13 lenses in their statistical sample selected from
8958 flat spectrum radio sources, plus 9 additional lenses
\citep{myers03,browne03} Thus, the final SQLS sample more than doubles
the number of quasar lenses in a single statistical lens sample, which
is a significant improvement. Furthermore, an advantage of SQLS over
CLASS is that the redshifts of all the source quasars are known, which
allows us to conduct more robust statistical analyses.  

It is expected that future wide-field optical imaging surveys such as
Pan-STARRS, Hyper Suprime-cam, the Dark Energy Survey, and the Large
Synoptic Survey Telescope will find many more strongly lensed quasars
\citep{oguri10}. These future surveys plan to explore the time domain,
which can make use of the time variability of quasar images to locate
lensed quasars efficiently \citep{kochanek06,lacki09}. In addition, the
ongoing SDSS-III survey \citep{aihara11,eisenstein11} plans to take
spectra of more than 100,000 quasars at $2.2<z<3$ selected from the
SDSS imaging data \citep{ross12}. We plan to apply the same algorithms
used in the SQLS to the SDSS-III data to discover many more lensed
quasars.

\acknowledgments

Use of the UH 2.2-m telescope and the UKIRT 3.8-m telescope for the
observations is supported by the National Astronomical
Observatory of Japan (NAOJ). 
Based in part on observations obtained with the Apache Point Observatory
3.5-meter telescope, which is owned and operated by the Astrophysical
Research Consortium. 
Based in part on data collected at Subaru Telescope, which is operated by the 
National Astronomical Observatory of Japan, and obtained from the SMOKA, 
which is operated by the Astronomy Data Center, National Astronomical 
Observatory of Japan.
Based in part on observations obtained at the Gemini Observatory, which is
operated by the Association of Universities for Research in Astronomy,
Inc., under a cooperative agreement with the NSF on behalf of the
Gemini partnership: the National Science Foundation (United States),
the Science and Technology Facilities Council (United Kingdom), the
National Research Council (Canada), CONICYT (Chile), the Australian
Research Council (Australia), Minist\'{e}rio da Ci\^{e}ncia e
Tecnologia (Brazil) and Ministerio de Ciencia, Tecnolog\'{i}a e
Innovaci\'{o}n Productiva (Argentina).
Telescopio Nazionale Galileo (TNG) operated on the island of La Palma by the 
Fundacion Galileo Galilei of the INAF (Istituto Nazionale di Astrofisica) at 
the Spanish Observatorio del Roque de los Muchachos of the Instituto de 
Astrofisica de Canarias.

This work was supported in part by the FIRST program "Subaru
Measurements of Images and Redshifts (SuMIRe)", World Premier
International Research Center Initiative (WPI Initiative), MEXT,
Japan, and Grant-in-Aid for Scientific Research from the JSPS
(23740161). 
This work is supported in part by JSPS Core-to-Core Program
``International Research Network for Dark Energy''.
N.~I. acknowledges support from MEXT KAKENHI 21740151.
M.~A.~S. acknowledges the support of NSF grant AST-0707266.
C.~E.~R. acknowledges the support of the JSPS Research Fellowship. 
C.~S.~K. is supported by NSF grant AST-1009756.
The Institute for Gravitation and the Cosmos is supported by
the Eberly College of Science and the Office of the Senior Vice
President for Research at the Pennsylvania State University.

Funding for the SDSS and SDSS-II has been provided by the Alfred
P. Sloan Foundation, the Participating Institutions, the National
Science Foundation, the U.S. Department of Energy, the National
Aeronautics and Space Administration, the Japanese Monbukagakusho, the
Max Planck Society, and the Higher Education  Funding Council for
England. The SDSS Web Site is http://www.sdss.org/. 

The SDSS is managed by the Astrophysical Research Consortium for the
Participating Institutions. The Participating Institutions are the
American Museum of Natural History, Astrophysical Institute Potsdam,
University of Basel, Cambridge University, Case Western Reserve
University, University of Chicago, Drexel University, Fermilab, the
Institute for Advanced Study, the Japan Participation Group, Johns
Hopkins University, the Joint Institute for Nuclear Astrophysics, the
Kavli Institute for Particle Astrophysics and Cosmology, the Korean
Scientist Group, the Chinese Academy of Sciences (LAMOST), Los Alamos
National Laboratory, the Max-Planck-Institute for Astronomy (MPIA),
the Max-Planck-Institute for Astrophysics (MPA), New Mexico State
University, Ohio State University, University of Pittsburgh,
University of Portsmouth, Princeton University, the United States
Naval Observatory, and the University of Washington.

\clearpage

\begin{deluxetable}{lcc}
\tabletypesize{\footnotesize}
\tablecaption{Source Quasars\label{tab:source}}
\tablewidth{0pt}
\tablehead{ 
\colhead{Name} &
\colhead{Redshift}    &
\colhead{$i$} 
}
\startdata
SDSSJ000009.26+151754.5   & 1.1985 & 19.057 \\
SDSSJ000009.42$-$102751.9 & 1.8449 & 18.698 \\
SDSSJ000013.14+141034.6   & 0.9582 & 19.028 \\
SDSSJ000017.38$-$085123.7 & 1.2491 & 18.204 \\
SDSSJ000024.02+152005.4   & 0.9889 & 18.969 \\
\enddata

\tablecomments{The redshift and $i$-band magnitude
(Galactic extinction corrected) distributions of the source
quasars are presented in their entirety in the electronic edition
of the {\it Astronomical Journal}. The first five rows are 
shown here for guidance regarding its form and content.
See also the SQLS webpage 
({\tt http://www-utap.phys.s.u-tokyo.ac.jp/\~{}sdss/sqls/}) for the
list of the 50,836 quasars.
}
 
\end{deluxetable}

\clearpage

\begin{deluxetable}{lr}
\tabletypesize{\footnotesize}
\tablecaption{Number of Candidates at Each Stage of the Selection 
Process\label{tab:process}}
\tablewidth{0pt}
\tablehead{ 
\colhead{Candidate Selection Process} & 
\colhead{Number} 
}
\startdata
\\
SDSS DR7 spectroscopically confirmed quasars                                                    & 105,783   \\
Quasars in the special plates                                                                   & $-$9,742  \\
Quasars {\it not} passing the criteria ($0.6<z<2.2$, $i<19.1$, and {\tt PSF\_WIDTH}$<1\farcs8$) & $-$45,205 \\ 
\\
Source quasars to complete the final statistical lens sample                                    & 50,836\tablenotemark{a} \\
\\
\hline
\\
\enddata

\tablenotetext{a}{Among the 50,836 quasars, 35,623 quasars were checked in 
Paper IV. The number of source quasars in Paper IV is 36,287, but 655
DR5 quasars are included in the special plates, or do not meet the
criteria to make our source quasar sample in the DR7 quasar catalog.}

\tablenotetext{b}{Seven candidates are selected by both the morphological 
and color selection algorithms, five of which were already studied in
Paper II and Paper IV. The remaining two candidates selected by both
the algorithms are listed in Table~\ref{tab:can_mor}.}
 
\end{deluxetable}

\clearpage

\begin{deluxetable}{lccccllll}
\tabletypesize{\footnotesize}
\rotate
\tablecaption{DR7 Morphologically Selected Candidates\label{tab:can_mor}}
\tablewidth{0pt}
\tablehead{ 
\colhead{Object} & 
\colhead{Redshift\tablenotemark{a}} &
\colhead{$i$\tablenotemark{b}} & 
\colhead{$\theta_{\rm SDSS}$\tablenotemark{c}} & 
\colhead{$|{\Delta} i|$\tablenotemark{c}} & 
\colhead{Image\tablenotemark{d}} & 
\colhead{Spec\tablenotemark{d}} & 
\colhead{Comment} & 
\colhead{Ref.} 
}

\startdata
SDSS~J074248.40+270412.2 & 0.626 & 19.05 & 0.78 & 1.74 & Te(I)                           & \nodata\phn   & single QSO                          & \nodata\phn \\
SDSS~J081405.56+582840.9 & 1.429 & 18.24 & 0.50 & 0.94 & Te(I)                           & \nodata\phn   & single QSO                          & \nodata\phn \\
SDSS~J090404.15+151254.5 & 1.825 & 17.50 & 1.04 & 1.64 & Te(V,R,I),Op(I)                 & FO            & SDSS lens                           & 1 \\
SDSS~J090645.62+122512.8 & 0.602 & 19.09 & 0.63 & 0.99 & Te(I)                           & \nodata\phn   & single QSO                          & \nodata\phn \\
SDSS~J092718.37+211357.4 & 1.845 & 17.24 & 2.36 & 0.21 & Te(I)                           & \nodata\phn   & no lensing object                   & \nodata\phn \\
SDSS~J093139.97+402548.4 & 1.028 & 18.35 & 0.43 & 0.21 & SP(i)                           & \nodata\phn   & single QSO                          & \nodata\phn \\
SDSS~J093527.90+232525.0 & 2.014 & 19.04 & 1.78 & 1.98 & Te(I)                           & \nodata\phn   & QSO+galaxy                          & \nodata\phn \\
SDSS~J095036.63+261706.1 & 1.264 & 18.60 & 0.44 & 0.96 & Te(I)                           & \nodata\phn   & single QSO                          & \nodata\phn \\
\enddata
\tablenotetext{a}{Redshifts from the SDSS DR7 quasar catalog.} 
\tablenotetext{b}{$i$-band PSF magnitudes with Galactic extinction
   corrections from the SDSS DR7 quasar catalog.}  
\tablenotetext{c}{Image separations ($\theta_{\rm SDSS}$) in units of
   arcsec and absolute $i$-band magnitude differences ($|{\Delta} i|$)
   between the expected two components, derived from fitting the SDSS
   $i$-band image of each candidate with two PSFs using GALFIT.} 
\tablenotetext{d}{Instruments (and filters) used for the follow-up
  observations.  
   Te: Tek2k CCD at UH88, Op: Optic CCD at UH88, 8k: UH8k at UH88, WF: WFGS2 at UH88,
   FO: FOCAS at Subaru, , SC: Suprime-Cam at Subaru, IR: IRCS at Subaru,
   SP: SPIcam at ARC 3.5m, NF: NICFPS at ARC 3.5m, 
   GM: GMOS at Gemini North,
   PA: PANIC at WB 6.5m,
   DO: DOLORES at TNG 3.6m.
   }
\tablerefs{(1) \citealt{kayo10}; 
           (2) \citealt{schechter98};
           (3) N. Inada, et al., in preparation;
           (4) \citealt{kayo07};
           (5) \citealt{oguri08d};
           (6) \citealt{jackson12}.
           }
\end{deluxetable}

\begin{deluxetable}{lcccllll}
\tabletypesize{\footnotesize}
\rotate
\tablecaption{DR7 Color Selected Candidates\label{tab:can_col}}
\tablewidth{0pt}
\tablehead{ 
\colhead{Object} & 
\colhead{Redshift\tablenotemark{a}} &
\colhead{$i$\tablenotemark{b}} & 
\colhead{$\theta_{\rm SDSS}$\tablenotemark{c}} & 
\colhead{Image\tablenotemark{d}} & 
\colhead{Spec\tablenotemark{d}} & 
\colhead{Comment} & 
\colhead{Ref.} 
}

\startdata
SDSS~J080713.57+155844.5 & 2.159       & 18.97 &       &                   &               &                          &  \\
SDSS~J080713.05+155838.5 & \nodata\phn & 19.70 &  9.72 & Te(I)             & \nodata\phn   & no lensing object        & \nodata\phn \\ \hline
SDSS~J080753.56+131505.7 & 1.131       & 18.27 &       &                   &               &                          &  \\
SDSS~J080753.74+131505.5 & \nodata\phn & 19.53 &  2.47 & Te(I)             & \nodata\phn   & QSO+galaxy               & \nodata\phn \\ \hline
SDSS~J081753.35+125821.5 & 1.975       & 18.77 &       &                   &               &                          &  \\
SDSS~J081754.04+125836.0 & 1.700       & 19.49 & 17.57 & \nodata\phn       & DA            & QSO pair                 & \nodata\phn \\ \hline
SDSS~J082234.33+170935.9 & 1.780       & 18.39 &       &                   &               &                          &  \\
SDSS~J082234.38+170955.0 & 1.798       & 19.65 & 19.02 & \nodata\phn       & DA            & QSO pair                 & \nodata\phn \\
\enddata
\tablenotetext{a}{Redshifts from the SDSS DR7 quasar catalog or
  follow-up spectroscopy.} 
\tablenotetext{b}{$i$-band PSF magnitudes with Galactic extinction
   corrections from the SDSS DR7 quasar catalog.}  
\tablenotetext{c}{Image separations ($\theta_{\rm SDSS}$) in units of
   arcsec between the two components from the SDSS imaging data.} 
\tablenotetext{d}{Instruments (and filters) used for the follow-up observations. 
   Te: Tek2k CCD at UH88, WF: WFGS2 at UH88, 8k: UH8k at UH88, 
   DA: DIS III at ARC 3.5m, NF: NICFPS at ARC 3.5m, SP: SPIcam at ARC 3.5m,
   DO: DOLORES at TNG 3.6m, 
   MS: MMT spectrograph, 
   FO: FOCAS at Subaru, MO: MOIRCS at Subaru
   GM: GMOS at Gemini North.
   }
\tablerefs{(1) C.~E. Rusu et al., in preparation; 
           (2) \citealt{myers08};
           (3) \citealt{myers07};
           (4) \citealt{kayo10}.
           }
\end{deluxetable}

\clearpage

\begin{deluxetable}{lccccrcccll}
\tabletypesize{\footnotesize}
\rotate
\tablecaption{DR7 Statistical Sample \label{tab:lens_dr7stat}}
\tablewidth{0pt}
\tablehead{ 
\colhead{Object} & 
\colhead{$N_{\rm img}$} & 
\colhead{$z_s$\tablenotemark{a}} &
\colhead{$z_l$\tablenotemark{b}} &
\colhead{$i$\tablenotemark{c}} &
\colhead{$\theta_{\rm max}$\tablenotemark{d}} & 
\colhead{$f_{I}$ \tablenotemark{e}} & 
\colhead{$I_{\rm gal}$ \tablenotemark{f}} & 
\colhead{Source} & 
\colhead{Non-SDSS designation} & 
\colhead{Ref.} 
}
\startdata
SDSS~J024634.09$-$082536.1 & 2 & 1.686 & 0.723       & 17.76 &  1.09 & 0.34 & 20.78(0.04) & DR3  &  \nodata\phn  & 1,2,3 \\ 
SDSS~J074653.04+440351.3   & 2 & 1.998 & 0.513       & 18.71 &  1.08 & 0.97 & 19.62(0.19) & DR5  &  \nodata\phn  & 4,5 \\ 
SDSS~J080623.70+200631.8   & 2 & 1.538 & 0.573       & 18.89 &  1.49 & 0.76 & 20.16(0.03) & DR5  &  \nodata\phn  & 2,6,7 \\ 
SDSS~J091301.02+525928.8   & 2 & 1.378 & (0.830)     & 16.17 &  1.11 & 0.70 & 18.81(0.45) & DR3  &  SBS~0909+523 & 3,8,9 \\
SDSS~J092455.79+021924.9   & 4 & 1.523 & 0.394       & 18.12 &  1.81 & 0.43 & 19.36(0.02) & DR3  &  \nodata\phn  & 3,9,10,11 \\ 
SDSS~J095122.57+263513.9   & 2 & 1.246 & 0.260       & 17.24 &  1.10 & 0.33 & 19.66(0.03) & DR7  &  FBQ0951+2635 & 2,3,12\\
SDSS~J100120.83+555349.6   & 2 & 1.413 & 0.36        & 16.68 &  6.17 & 0.94 & 17.11(0.02) & DR3  &  Q0957+561    & 3,13,14 \\
SDSS~J100128.61+502756.8   & 2 & 1.841 & 0.415       & 17.32 &  2.86 & 0.86 & 19.63(0.03) & DR3  &  \nodata\phn  & 15  \\
\enddata
\end{deluxetable}

\begin{deluxetable}{lccccrccclll}
\tabletypesize{\scriptsize}
\rotate
\tablecaption{DR7 Additional Lensed Quasars\label{tab:lens_dr7add}}
\tablewidth{0pt}
\tablehead{ 
\colhead{Object} & 
\colhead{$N_{\rm img}$} & 
\colhead{$z_s$\tablenotemark{a}} &
\colhead{$z_l$\tablenotemark{b}} &
\colhead{$i$\tablenotemark{c}} &
\colhead{$\theta$\tablenotemark{d}} & 
\colhead{$f_{I}$ \tablenotemark{e}} & 
\colhead{$I_{\rm gal}$ \tablenotemark{f}} & 
\colhead{Source} & 
\colhead{Non-SDSS designation} & 
\colhead{Reject\tablenotemark{g}} & 
\colhead{Ref.} 
}
\startdata
SDSS~J013435.66$-$093102.9 & 5 & 2.220 & 0.765       & 19.62 &  0.68 & 0.03 & 19.27(0.08) & DR3 & PMN~J0134$-$0931  & $z>2.2$, $i>19.1$, $\theta<1''$   & 1,2,3,4 \\  
SDSS~J014516.59$-$094517.3 & 2 & 2.730 & 0.491       & 16.57 &  2.23 & 0.15 & 18.72(0.03) & DR3 & Q0142$-$100       & $z>2.2$, $|{\Delta} I|>1.25$      & 4,5,6 \\ 
SDSS~J074352.61+245743.6   & 2 & 2.165 & 0.381       & 19.01 &  1.03 & 0.26 & 20.47(0.37) & DR3 & ULAS~J0743+2457   & $|{\Delta} I|>1.25$               & 7,8 \\
SDSS~J081331.28+254503.0   & 4 & 1.510 & $\sim 0.1$  & 15.30 &  0.91 & 0.06 & 16.62(0.81) & DR3 & HS0810+2554       & $\theta<1''$                      & 4,9 \\
SDSS~J081959.80+535624.2   & 2 & 2.239 & 0.294       & 18.52 &  4.04 & 0.23 & 16.54(0.04) & DR5 & \nodata\phn       & $z>2.2$, $|{\Delta} I|>1.25$      & 10  \\
SDSS~J082016.11+081215.9   & 2 & 2.024 & 0.803       & 19.05 &  2.20 & 0.12 & 19.59(0.06) & DR5 & ULAS~J0820+0812   & $|{\Delta} I|>1.25$               & 11 \\ 
SDSS~J082138.94+121729.9   & 2 & 3.113 & 0.39        & 18.76 &  2.56 & 0.14 & 19.05(0.22) & DR7 & HS0818+1227       & $z>2.2$, $|{\Delta} I|>1.25$      & 4,12\\
SDSS~J083216.99+040405.2   & 2 & 1.116 & 0.659       & 18.89 &  1.98 & 0.22 & 19.35(0.03) & DR3 & \nodata\phn       & $|{\Delta} I|>1.25$               & 13 \\
\enddata
\end{deluxetable}

\clearpage

\clearpage

\end{document}